\documentclass[aps,prb,twocolumn,floatfix,showpacs]{revtex4-1}
\usepackage{amsmath,amssymb,graphicx,bm} 
\usepackage[xcolor=pst]{pstricks}

\begin{document} \title{Mean-field solution of the Potts glass near the transition temperature to the ordered phase}

\author{V.  Jani\v{s}} \author{A. Kl\'\i\v{c}}

\affiliation{Institute of Physics, Academy of Sciences of the Czech
  Republic, Na Slovance 2, CZ-18221 Praha, Czech Republic }
\email{janis@fzu.cz}

\date{\today}


\begin{abstract}
We expand asymptotically mean-field solutions of the $p<4$ Potts glass with various levels of replica-symmetry breaking  below the transition temperature to the glassy phase. We find that the ordered phase is degenerate and solutions with one hierarchy of spin replicas and with the full continuous replica-symmetry breaking coexist for $p> p^{*} \approx 2.82$. The latter emerges immediately with the instability of the replica-symmetric one. Apart from these two solutions there exists also a succession of unstable states converging to the solution with the continuous replica-symmetry breaking that is marginally stable and has the highest free energy.   
\end{abstract}
\pacs{64.60.Cn,75.50.Lk}

\maketitle 

\section{Introduction}
\label{sec:Intro}

Models of spin glasses have been attracting a considerable attention of condensed matter theorists for more than three decades. Original motivation for constructing microscopic models of spin glasses came from experimental observation of an unusual low-temperature behavior of magnetic impurities randomly diluted in a nonmagnetic metal.\cite{Cannella72} It soon became clear that a new type of long-range order of magnetic impurities emerges in spin glasses, qualitative understanding of which demands theoretical modeling. Rather soon Edwards and Anderson proposed a generic model of Ising spins with a randomly distributed exchange interaction.\cite{Edwards75}  A naive mean-field solution,\cite{Sherrington75} now known as the Sherrington-Kirkpatrick (SK) model, started a wave of intense investigation of Ising and Heisenberg spin glasses. The reason for the extended interest in spin-glass theories was insufficient understanding of the inconsistency of the SK mean-field approximation.  A fully physically consistent solution was then proposed by Parisi via the replica trick used to handle quenched averaging over random configurations of the spin exchange.\cite{Parisi80} A rigorous proof of exactness of the Parisi free energy of the SK model, known as a full replica-symmetry breaking (FRSB), was completed only recently.\cite{Guerra03,Talagrand06}          

The full mean-field solution of the Ising spin glass is not only complicated in its analytic structure but also in its physical interpretation.\cite{Mezard87} That is why simpler models, random energy,\cite{Derrida81} $p$-spin,\cite{Gardner85} Potts,\cite{Elderfield83} or quadrupolar\cite{Goldbart85b} glass models,  have been introduced so that the origin and the meaning of replica-symmetry breaking in frustrated spin models can be better understood. All four models behave differently from the Heisenberg spin glasses. They show temperature intervals where a first step toward the Parisi solution in the replica trick, the so-called one-level replica symmetry breaking (1RSB), is locally stable. While 1RSB is the true equilibrium state  of the first model in low temperatures, it is locally stable only in a small interval of temperatures close to the transition temperature to the glassy phase in the latter three. A full replica-symmetry breaking solution with infinite-many hierarchies of replicated spin variables should lead to a marginally stable equilibrium state at very low temperatures there. The genesis of FRSB and the way a  solution with a continuous distribution of hierarchies of replicated spin variables is reached in these models without reflection symmetry in spin variables has not yet been fully clarified.        

We recently demonstrated that, although the first level of replica-symmetry breaking below the transition temperature to the glassy phase in the mean-field model of the Potts glass is locally stable, a full continuous replica-symmetry breaking solution coexist with it and has a higher free energy.\cite{Janis11a}  This result suggests that FRSB leads even in models without reflection symmetry to the true equilibrium state independently of the fact that a replica-symmetry breaking solution with a finite number of replica hierarchies is locally stable.  It has been assumed that the Parisi continuous replica-symmetry breaking in models without spin-reflection symmetry fails or is inconsistent.\cite{Goldbart85a,Gross85} A natural question arises when a locally stable discrete and marginally stable continuous RSB solutions coexist: How does the FRSB solution emerge when a discrete RSB solution with finite-many replica hierarchies no longer decays to solution with a higher number of replica hierarchies. The full RSB state has been assumed to emerge only below the temperature $T_{2}$ at which the 1RSB solution becomes again unstable.  Heuristic arguments were used to suggest a cascade of successive transitions below the instability of 1RSB.\cite{Gribova03,Schelkacheva10}    

The aim of this paper is to demonstrate explicitly the genesis of the Parisi solution with a continuous FRSB in the model of the Potts glass with $p$ states. We use an asymptotic expansion of the Parisi solution near the critical temperature and show that for $p\le 4$ the solution with FRSB emerges at the transition point at which the replica-symmetric solution gets unstable and the system undergoes a continuous transition to the glassy phase.  We find that near the transition temperature to the glassy phase there exists apart from a locally stable 1RSB solution for $p^{*}\le p < 4$ also a series of unstable solutions with $K=1,2,\ldots$ discrete hierarchies  breaking replica symmetry and converging towards a marginally stable Parisi solution with a continuous FRSB.

\section{Mean-field model}
\label{sec:MFM}

\subsection{Potts Hamiltonian and a replicated mean-field solution}
\label{sec:MFM-HR}

The Potts model is a generalization of the Ising model to more than
two spin components.  The original formulation of Potts \cite{Potts52} with
 Hamiltonian $H_{p}=-\sum_{i<j}J_{ij}\delta_{n_{i},n_{j}}$ where
$n_{i}=0,\ldots, p-1$ is an admissible value of the $p$-component
model on the lattice site $\mathbf{R}_{i}$, is unsuitable for
practical calculations. The Potts Hamiltonian can, however, be represented via
interacting spins \cite{Wu82} %
\begin{equation}\label{eq:Hamiltonian}
  H_{p}=-\frac{1}{2}\sum_{i,j}J_{ij}\mathbf{S}_{i}\cdot \mathbf{S}_{j}  - \sum_{i}\mathbf{h}\cdot \mathbf{S}_{i}\ ,
\end{equation} %
where $\mathbf{S}_{i} = \{s_{i}^{1},\ldots s_{i}^{p-1} \}$ are Potts
vector variables taking values from a set of state vectors
$\{\mathbf{e}_{A}\}_{A=1}^{p}$. Functions on vectors $\mathbf{e}_{A}$
are in equilibrium fully defined through their scalar product
\begin{subequations}
\begin{equation}\label{eq:Potts-Scalar}
e^{\alpha}_{A}e^{\alpha}_{B}=p\ \delta_{A B}-1 ,
\end{equation}
where $\alpha\in\{1,...,p-1\}$. We use the Einstein summation convention for
repeating Greek indices of the vector components indicating a scalar
product of the Potts vectors. 

The components of the Potts vectors obey the following sum rules
\begin{align}\label{eq:Potts-restrict}
\sum_{A=1}^{p}e^{\alpha}_{A}&=0 \\ \label{eq:Potts-norm}
\sum_{A=1}^{p}e^{\alpha}_{A}e^{\beta}_{A}&=p\ \delta^{\alpha \beta} 
\end{align}
\end{subequations}
from which we can construct their explicit representation 
\begin{align}\label{eq:Potts-repre}
e^{\alpha}_{A}=\left\{
\begin{array}{ll}
   0 & A<\alpha \\
   \sqrt{\frac{p(p-\alpha)}{p+1-\alpha}}  & A=\alpha \\
   \frac{1}{\alpha-p}\sqrt{\frac{p(p-\alpha)}{p+1-\alpha}} &
   A>\alpha\ .
\end{array}
\right. 
\end{align}
It is evident from the representation of Potts vectors in Eq.~\eqref{eq:Potts-repre} that the Potts variables are not symmetric around zero for $p > 2$, that is the Potts model does not possess spin-reflection symmetry. In case of $p=2$ the Potts model reduces to the Ising model. 

Frustration due to a quenched randomness in the Potts model is introduced via static fluctuations of the interaction parameters $J_{ij}$ being distributed randomly with probability 
\begin{equation}\label{eq:distributionJ}
  P(J_{ij})=\frac{1}{(2\pi
  J^{2}/N)^{1/2}}\exp{\frac{-(J_{ij}-J_{0})^{2}}{2J^{2}/N}}\ ,
\end{equation}
where $J_{0}=N^{-1}\sum_{j} J_{0j}$ is the averaged (ferromagnetic) interaction and $N$ is the number of lattice sites. Randomness is assumed to be  quenched, that is, the equilibrium free energy is averaged over the configurations of the spin-spin interaction. The spin-spin interaction is long-range with an infinitesimal ($N^{-1}$) variance of its fluctuations (mean-field model).  

The standard way to derive a mean-field approximation to frustrated models with random exchange interactions is to use the replica trick with which a quenched averaging is transformed to an annealed one of a replicated system.\cite{Mezard87} There is, however, a price we pay for this simplifying transformation.  We must perform a limit to zero number of replicas, which demands analytic continuation of the solution in the replica number. This is possible only under specific conditions with an appropriate symmetry of the order parameters in the replicated phase space.\cite{Parisi80}  There is a possibility to avoid the replica trick with the limit to zero replicas in that we demand thermodynamic homogeneity of the averaged free energy.\cite{Janis05}  Although replicas are used, there is no need for the limit to zero number of replicas. Instead, independence of the replication index is demanded in the real-replica approach.  The replicated mean-field free energy of the Potts glass with the gaussian distribution of spin-spin interactions and the probability distribution from Eq.~\eqref{eq:distributionJ} is
\begin{multline}\label{eq:FE-averaged-finite}
f_\nu = \frac{\beta J^2}{4} \left\{\frac 1\nu\sum_{a\neq b}^\nu
\chi^{\alpha\beta}_{ab}\left[\chi^{\alpha\beta}_{ab} + 2 q\delta^{\alpha\beta}\right] \right. \\ \left. - (p- 1) (1 -
q)^2\right. \bigg\}
- \frac{\overline{J}_{0}}{2\nu}\sum_{a}^{\nu}m^{\alpha}_{a} m^{\alpha}_{a}
 -\frac 1{\beta\nu}\!\int\limits_{-\infty}^{\infty}\mathcal{D}_{(p-1)}(\mathbf{y}) \\
 \ln \text{Tr}_{\mathbf{S}} \exp\left\{\beta^2J^2\sum_{a < b}^\nu
\chi^{\alpha\beta}_{ab}S^{\alpha}_{a}S^{\beta}_{b} + \beta \sum_{a=1}^\nu \overline{h}^{\alpha}_{a}S^{\alpha}_{a}\right\}
\end{multline}
where $\nu$ is the number of replicas, 
$\overline{h}^{\alpha}_{a} = h^{\alpha} + \overline{J}_{0}m^{\alpha}_{a } + J y^{\alpha} \sqrt{q} $ is an effective magnetic field, and $\eta^{\alpha}$ are Gaussian random fields with a $(p-1)$-dimensional measure 
$$
\mathcal{D}_{(p-1)}(\mathbf{y})= \prod_{\alpha=1}^{p-1} \frac{d\eta^{\alpha}}{\sqrt{2\pi}} \exp\left\{- \frac{\left(y^{\alpha}\right)^{2}}2\right\} \ .
$$

We further denoted an effective ''ferromagnetic'' exchange 
$\overline{J}_{0}= J_{0} + \beta J^{2}(p - 2)/2$. The order parameters in the replicated free energy $f_{\nu}$ are the averaged square local magnetization $q = N^{-1}\sum_{i}m^{\alpha}_{i}m^{\alpha}_{i}$ and local overlap susceptibilities $\chi^{\alpha\beta}_{ab}= N^{-1}\sum_{i}\left[S^{\alpha}_{i,a}S^{\beta}_{i,b} - m^{\alpha}_{i,a}m^{\beta}_{i,b}\right]$ for  $a\neq b$ measuring the linear response of the replicated system to a small inter-replica interaction. If the free energy is thermodynamically homogeneous, the overlap susceptibilities must vanish in equilibrium, a saddle point of the replicated free energy. Free energy is then independent of the replica index $\nu$. It reduces in this case  to the replica-symmetric one having a representation for the isotropic Potts model (zero magnetic field and no long-range ferromagnetic order, $m^{\alpha}_{i}=0$)
\begin{subequations}\label{eq:FE-RS}
\begin{equation}\label{eq:FE-f_nu}
  -\beta f_{RS}=\frac{\beta^{2}}{4}(p-1)(q-1)^{2}+\int \mathcal{D}_{(p-1)} (\mathbf{y})\ln{Z_{0}}(\mathbf{y})\ .
\end{equation}
We denoted the local partition function of the Potts model in a random magnetic field 
\begin{equation}\label{eq:FE-Z0}
  Z_{0}(\mathbf{y})=\sum_{A=1}^{p}{\exp\{\beta \sqrt{q}y^{\alpha} e^{\alpha}_{A}\}}\ .
\end{equation}
\end{subequations}

We need to break the replica symmetry so that to check thermodynamic homogeneity of the equilibrium free energy from Eq.~\eqref{eq:FE-RS}. That is, to test stability of the replica-symmetric solution with respect to replications of the phase space of the relevant order parameters.

\subsection{Free energy with discrete hierarchies of replica-symmetry breaking}
\label{sec:MFM-Discrete}

A natural way to start with replications of the original model is to use simply two replicas. It was done in  detail for the Ising spin glass.\cite{Janis06a} It was demonstrated there that although free energy was lowered in the system with two replicas the instability and thermodynamic inhomogeneity of the replica-symmetric solution was made worse. It is necessary to continue analytically the replicated free energy to an arbitrary positive replication index to analyze dependence of free energy on the replication index. Making the replication index a continuous variable is possible only if the symmetry of the matrix of the local overlap susceptibilities possesses the symmetry introduced by Parisi in his construction within the replica trick.  Instead of numbers of replicas used we introduce a number of hierarchies of replicas that distinguish different solutions. A solution with $K$ hierarchies is determined from a saddle point of a free-energy functional
\begin{subequations}\label{eq:FE-K_general} 
  \begin{multline}\label{eq:FE-K} -\beta f_{K}(q,\{\Delta\chi_{l}\},
    \{ m_{l}\}) =\frac{\beta^{2}J^{2}}{4}(p-1)\bigg\{\bigg(1-q \\ -
    \sum_{j=1}^{K}\Delta \chi_{j}\Bigg)^{2} -\sum_{j=1}^{K}m_{j}\Delta
    \chi_{j}\bigg[\Delta \chi_{j}\\ +2\bigg(q+\sum_{l=j+1}^{K}\Delta
    \chi_{l}\bigg)\bigg]\Bigg\} +\int \mathcal{D}_{(p-1)}(\mathbf{y})
    \ln{Z_{K}^{K}}(\mathbf{y}) \end{multline}
  where%
  \begin{equation}\label{eq:zK}
    Z_{l}^{K}\left( \mathbf{y},\{\mathbf{\lambda}\}_{l+1}\right )
    =\left[\int
      \mathcal{D}_{(p-1)}
      (\mathbf{\lambda}_{l})\left(Z_{l-1}^{K}\right)\left(\mathbf{y},\{\mathbf{\lambda}\}_l\right)^{m_{l}}
    \right]^{\frac{1}{m_{l}}}\    
  \end{equation}%
  and $\{\lambda\}_{l}=\lambda_{l}, \ldots,\lambda_{K}$.  The initial zero-level partial sum reads
  \begin{multline}\label{eq:zK0} Z_{0}\left(\mathbf{y},\mathbf{ \lambda}\right) \equiv
    Z_{0}^{K}\left(\mathbf{y},\{ \mathbf{\lambda}\}_{1}\right)
    \\=\sum_{A=1}^{p}\exp\left\{\beta
      J\left(\sqrt{q}y^{\alpha}+\sum_{j=1}^{K}\sqrt{\Delta
          \chi_{j}}\lambda_{j}^{\alpha}\right)e^{\alpha}_{A}\right\}\    .
  \end{multline}\end{subequations} %
The equilibrium state for this free energy is characterized by the averaged square magnetization $q$ and a set of $K$ pairs $\Delta\chi_{l}, m_{l}$ for $l=1,2,\ldots, K$. Thermodynamic homogeneity of a $K$-level hierarchical solution is achieved if $\Delta\chi_{K+1} =0$, which leads to independence of the system on the next replicating parameter  $m_{K+1}$.\cite{Janis05}  The overlap susceptibilities $0 \le \Delta\chi_{l}\le 1$ form generally a decreasing sequence, since we demand that the last one should vanish in the thermodynamically homogeneous system. The indices counting the replica hierarchies $m_{l}$ can be arbitrary. There is, however, a degeneracy in the hierarchical free energy. We obtain $f_{K+1} = f_{K}$ if $m_{K+1} = 0, m_{k}, \infty$.  It means that  we obtain at least one new equilibrium solution for a pair $\Delta\chi_{K+1}, m_{K+1}$ with $m_{K+1}< m_{K}$. The new mean-field solution is acceptable if it leads to a smaller thermodynamic inhomogeneity measured by $\Delta\chi_{K+1} < \Delta\chi_{K}$. In the asymptotic solution near the critical temperature of the  Ising spin glass  the new solution with $m_{K+1}< m_{K}$ leads to a higher free energy $f_{K+1} > f_{K}$.\cite{Janis06c,Janis08b}  There is also another stationary solution for $m_{K+1} > m_{K}$ that, in the Ising model, lowers the free energy and worsens thermodynamic inhomogeneity. Hence, it is unacceptable.  We can do a similar analysis of the $K$RSB free energy, Eq.~\eqref{eq:FE-K_general},  near the transition temperature of the Potts glass.

We start with a solution with a first level of replica-symmetry breaking. An explicit mean-field free energy of the Potts glass with 1RSB reads  
\begin{multline}\label{eq:FE-1RSB}
-\beta f_{1}= \frac{\beta^{2}}4 (p-1)\\ \times \left[\left( 1 - q -\Delta\chi\right)^{2} - m\Delta\chi \left(\Delta\chi + 2 q\right) \right] \\
 + \frac 1m \int \mathcal{D}_{(p-1)} (\mathbf{y})\ \ln \int  \mathcal{D}_{(p-1)}(\mathbf{\lambda})\  Z_{0}^{(p)}\left(\mathbf{y},\mathbf{\lambda}\right)^{m}\ .   
\end{multline}  
The partition function for the $p$-state Potts model is constructed by using the representation in Eq.~\eqref{eq:Potts-repre}. For the three-state model we obtain explicitly
 \begin{multline}\label{eq:z30} 
  Z^{(3)}_{0}\left(\mathbf{y},\{\mathbf{\lambda}\}_{1}\right) = \exp\left\{\beta
      J\sqrt{2} \left(\sqrt{q}y^{1}+\sum_{j=1}^{K}\sqrt{\Delta
          \chi_{j}}\lambda_{j}^{1}\right)\right\}\\  +  \exp\left\{\beta
      J\left[\sqrt{\frac 32}\left(\sqrt{q}y^{2}+\sum_{j=1}^{K}\sqrt{\Delta
          \chi_{j}}\lambda_{j}^{2}\right) \right.\right. \\ \left.\left. - \sqrt{\frac 12}\left(\sqrt{q}y^{1}+\sum_{j=1}^{K}\sqrt{\Delta \chi_{j}}\lambda_{j}^{1}\right)\right]\right\} \\ 
 +  \exp\left\{- \beta
      J\left[\sqrt{\frac 32}\left(\sqrt{q}y^{2}+\sum_{j=1}^{K}\sqrt{\Delta
          \chi_{j}}\lambda_{j}^{2}\right) \right.\right. \\ \left.\left. + \sqrt{\frac 12}\left(\sqrt{q}y^{1}+\sum_{j=1}^{K}\sqrt{\Delta \chi_{j}}\lambda_{j}^{1}\right)\right]\right\} \ . 
  \end{multline}
Properties of the mean-field theory of the Potts glass with 1RSB have been analyzed by several groups.\cite{Cwilich89,Santis95,Gribova03} The most prominent feature of the solution of the Potts mean-field model with 1RSB and the number of states $p \ge 3$ is  its local stability near the transition to the low-temperature glassy phase.  One could conclude from this result that the hierarchical construction of the mean-field free energy stops just at 1RSB and no Parisi solution with FRSB exists in the region of local stability of 1RSB. We demonstrate on the asymptotic solution below the transition temperature to the glassy phase that the mean-field equations  of the Potts glass are degenerate and allow for a cascade of coexisting metastable sates including the Parisi FRSB solution that is marginally stable alike in the Sherrington-Kirpatrick model.

\subsection{Parisi solution with a continuous replica-symmetry breaking}
\label{sec:MFM-Parisi}

It is not necessary to derive a mean-field free energy with FRSB via checking stability of free energies with finite-many hierarchies of discrete replica symmetry breakings. It is sufficient to look at the behavior of the hierarchical free energy in the continuous limit. It obeys a differential equation derived first by Parisi for the Ising spin glass.\cite{Parisi80}  If we introduce a parameter $\lambda \in (0,1)$ and denote $X= q_{EA} - q= \sum_{l} \Delta\chi_{l}$, then the $\lambda$-dependent freee energy $g(\lambda,h)$  in the continuous limit of the replica symmetry breaking hierarchy with $K\to \infty$ and $\Delta\chi_{l}\propto K^{-1}$  obeys a differential equation that for a many-component spin model reads  
\begin{align}\label{eq:g0-DE}
  \frac{\partial g(\lambda,\mathbf{h})}{\partial \lambda} &= \frac {X}{2}
  \left[\frac{\partial^2 g(\lambda,\mathbf{h})}{\partial h^{\alpha}\partial h^{\alpha} } + m(\lambda) \frac{\partial g(\lambda,\mathbf{h})} {\partial h^{\alpha}}\frac{\partial g(\lambda,\mathbf{h})} {\partial h^{\alpha}} \right] 
\end{align}
where $m(\lambda)$ is a continuous limit of the replica indices $m_{l}$ from the solutions with discrete replica hierarchies. 

Having this differential equation we can try to resolve it on a phase space of the order-parameter functions $m(\lambda)$.  One of us recently suggested an explicit representation  for the free energy obeying the Parisi differential equation \eqref{eq:g0-DE}.\cite{Janis08a} It can easily be generalized also to the Potts glass for which we obtain
\begin{subequations}\label{eq:FE-continuous}
  \begin{multline}\label{eq:freeEcontinuous} -\beta
    f_{c}[q,X,m(\lambda)]=\log p+\frac{\beta ^{2}}{4}(p-1)(1-q-X)^{2} \\
    -\frac{\beta ^{2}}{2}(p-1)X \int_{0}^{1}d\lambda\
    m(\lambda)[q+X(1-\lambda)]+\left\langle g(1,\mathbf{y}
      \sqrt{q})\right\rangle_{y} \end{multline} %
  where $\langle F(\mathbf{y})\rangle_{\mathbf{y}} = \int \mathcal{D}_{(p-1)}(\mathbf{y}) F(\mathbf{y})$ and%
  \begin{multline}\label{eq:E0} g(\nu,\textbf{h})
    =\mathbb{T}_{\lambda} \exp\bigg\{\frac{X}{2}\int_0^{\nu} d\lambda
    \left[
      \partial_{\bar{h}^{\alpha}}\partial_{\bar{h}^{\alpha}}
    \right. \\ \left.+ m(\lambda) g^{\prime}_{\alpha}(\lambda,
      \mathbf{h} + \bar{\mathbf{h}})\partial_{\bar{h}^{\alpha}}
    \right] \bigg\} g_{0}(\mathbf{h}+
    \mathbf{\bar{h}})\big|_{\bar{\mathbf{h}} = 0} 
  \end{multline}\end{subequations}
with the initial local free energy  $ g_{0}(\mathbf{h})=\ln \sum_{A=1}^{p}\exp\{\beta h^{\alpha}e^{\alpha}_{A}\}$. We introduced an evolution operator represented via a
``time-ordering'' operator $\mathbb T_{\lambda}$ ordering products of
$\lambda$-dependent non-commuting operators from left to right in
$\lambda$-decreasing succession. We further denoted
$g^{\prime}_{\alpha}(\lambda, \mathbf{h})
\equiv \partial_{h_{\alpha}}g (\lambda, \mathbf{h}) $. We used an
auxiliary vector field $\bar{\mathbf{h}} = (\bar{h}^{1},\bar{h}^{2},\ldots \bar{h}^{p-1})$ to generate the necessary derivatives of the bare free energy $g_{0}$.

Free energy, Eq.~\eqref{eq:FE-continuous}, is stationary with respect to variations of variables $q$ and $X$ and function $m(\lambda)$ for every $\lambda \in (0,1)$. It is straightforward to show that $g(\lambda,\mathbf{h})$ obeys differential equation~\eqref{eq:g0-DE}.  Although representation~\eqref{eq:FE-continuous} of a mean-field free energy with a continuous FRSB of the Potts glass is implicit, it is self-contained and allows us to derive explicit stationarity equations  for the order parameters and other quantities. It also enables to reach  directly approximate or asymptotic solutions without using solutions with finite numbers of replica hierarchies and studying their stability. We use representation \eqref{eq:FE-continuous}  here to derive an asymptotic solution with a continuous replica-symmetry breaking of the Potts glass with $p\le 4$ states below the transition temperature to the glassy phase.

\section{Asymptotic solution near the critical temperature}
\label{sec:AS}

Full-scale solutions of the mean-field equations of the Potts glass are not available due to their complex structure. What we can explicitly obtain are only asymptotic  limits  
of the solutions with various degrees of replica-symmetry breaking. We hence expand asymptotically the replica-symmetric solution, the solutions with discrete finite-many hierarchies of the replicated spin variables as well as  the Parisi solution with the continuous order-parameter function. The expansion coefficients will be calculated with program MATHEMATICA. We expand the corresponding free energies to the fifth order in the small expansion parameter being $\tau= (1 - T/T_{c})$, where $T_{c} = J = 1$ is the temperature of a continuous transition to the low-temperature glassy phase at which different mean-solutions can be distinguished. Since we assume a continuous transition to the glassy phase, our analysis restricts to the Potts model with the number of states $p\le 4$. The derived asymptotic solutions allow us, however, to analyze the asymptotic behavior of the glassy phase below the transition temperature as a function of a continuous parameter $p$.   

\subsection{Replica-symmetric solution}
\label{sec:AS-RS}

The simplest mean-field state is a stationary solution of the replica-symmetric free energy from Eq.~\eqref{eq:FE-RS}. The only order parameter is the average of the square of local magnetizations $q = N^{-1}\sum_{i}m_{i}^{\alpha}m_{i}^{\alpha} = \langle m^{\alpha} m^{\alpha}\rangle_{\mathbf{y}}$.   This parameter vanishes in the high-temperature paramagnetic phase and starts to grow continuously from zero below the transition point to the glassy phase.  The corresponding stationarity equation for the replica-symmetric order parameter derived from free energy, Eq.~\eqref{eq:FE-RS}, reads 
\begin{equation}\label{eq:RS-q}
  (p-1)q+1=
  p \sum_{A=1}^{p}\int \mathcal{D}_{(p-1)}(\mathbf{y}) \frac{\exp\{2\beta 
  \sqrt{q}y^{\alpha}e^{\alpha}_{A}\}}{Z_{0}(\mathbf{y})^{2}}\ .
\end{equation}
 We do not want to evaluate fully the gaussian integrals on the right-hand side of Eq.~\eqref{eq:RS-q} but rather only near the transition temperature. Since the transition is expected to be continuous we can assume that the order parameter $q$ is small and expand free energy, Eq.~\eqref{eq:FE-RS},  into a power series in $q$. We cut the expansion at the fifth order. The explicit expression for the expanded free energy is given in Appendix in Eq.~\eqref{eq:FE-RS_beta}. The first term in the expansion of the free energy proportional to $q^{2}$ changes sign at $\beta = 1$ indicating a continuous transition to an ordered phase. The transition is continuous where the third order of the expansion is positive,  that is, for $p\le 6$. 

It is sufficient to expand order parameter $q$ to the third order of the the expansion parameter $\tau = 1 - T$ so that to obtain the expansion of free energy to the fifth order. The expansion calculated with program MATHEMATICA reads  
\begin{multline}\label{eq:q-RS:tau}
q \doteq \frac{4 \tau }{6-p} +\frac{2 \left(-7 p^2-60 p+180\right) \tau
^2}{3 (6-p)^3} \\+\frac{8 \left(p^4+300
   p^3-1044 p^2+4320 p-7776\right) \tau ^3}{9 (6-p)^5}\ .
\end{multline}
Inserting this expansion into Eq.~\eqref{eq:FE-RS_beta} we obtain an asymptotic expression for the replica-symmetric free energy 
\begin{multline}\label{eq:FE-RS_tau}
\frac{\beta}{p-1} f_{RS}\doteq\frac{8 \tau ^3}{3
   (6-p)^2}+
\frac{4 \left(p^2-84 p+252\right) \tau ^4}{3
(6-p)^4}\\+\frac{2 \left(29 p^4+1320
   p^3+54360 p^2-294624 p+421200\right) \tau ^5}{45 (6-p)^6}\ .
\end{multline}
The asymptotic expansion based on smallness of the order parameter breaks down at $p=6$ indicating a change in the way the transition to the glassy phase occurs. Notice that the expanded free energy of the Potts glass in Eq.~\eqref{eq:FE-RS_tau} coincides with that of the Ising spin glass for $p=2$.\cite{Janis08b} 

\subsection{1RSB solution}
\label{sec:AS-1RSB}

The next step beyond the replica-symmetric solution is a state with the first level of replica-symmetry breaking. Its free energy is described by three order parameters $q,\Delta\chi, m$. We derive their defining equations from the stationarity point of the 1RSB free energy in Eq.~\eqref{eq:FE-1RSB}.  To be able to write down these equations in a condensed way we first introduce a useful notation. 
\begin{subequations}
\begin{align}\label{eq:1RSB-zA}
Z_{0}(\mathbf{y},\mathbf{\lambda}) & = \sum_{A}E_{A}(\mathbf{y},\mathbf{\lambda})\ ,\\
E_{A}(\mathbf{\lambda},\mathbf{y})& = \exp\left\{\beta\left[\sqrt{q} y^{\alpha} + \sqrt{\Delta\chi} \lambda^{\alpha}\right]e^{\alpha}_{A}\right\}\ . 
\end{align}
Further on we will need
\begin{align}\label{eq:1RSB-t_alpha}
t^{\alpha}& = \frac{\sum_{A}e^{\alpha}_{A}E_{A}(\mathbf{y},\mathbf{\lambda}) }{Z_{0}(\mathbf{y},\mathbf{\lambda}) }\ ,
\\
\rho & = \frac{Z_{0}(\mathbf{y},\mathbf{\lambda})^{m}}{\langle Z_{0}(\mathbf{y},\mathbf{\lambda})^{m}\rangle_{\lambda}} \ .
\end{align}
\end{subequations}

An equation for the equilibrium order parameter $q$ derived from $\partial f_{1}/\partial q = 0$ reads
\begin{multline}\label{eq:1RSB-q}
(p-1)(q+(1-m)\Delta\chi)=(1-m)\langle\langle \rho\ t^{\alpha}
t^{\alpha}\rangle_{\lambda} \rangle_{y}\\ +m\langle\langle\rho\ 
t^{\alpha}\rangle_{\lambda} \langle\rho\ t^{\alpha}\rangle_{\lambda}
\rangle_{y}\ .
\end{multline}
Analogously from $\partial f_{1}/\partial \Delta\chi = 0$ we obtain 
\begin{equation}\label{eq:1RSB-Dchi}
(p-1)(q+\Delta\chi)=\langle\langle \rho\ t^{\alpha}
t^{\alpha}\rangle_{\lambda} \rangle_{y}.
\end{equation}
Finally the equation for parameter $m$ is
\begin{multline}\label{eq:1RSB-m}
m\frac{\beta^{2}}{4}(p-1)\Delta\chi(2q+\Delta\chi)=\langle\langle \rho \ln
Z_{0}(\mathbf{y},\mathbf{\lambda}) \rangle_{\lambda} \rangle_{y}\\ - \langle \ln Z_{1}(\mathbf{y})  \rangle_{y}
\end{multline}
where $Z_{1}(\mathbf{y}) = \left[\int \mathcal{D}_{(p-1)}(\mathcal{\lambda}) Z_{0}(\mathbf{y},\mathbf{\lambda})^{m}\right]^{1/m}\equiv \langle Z_{0}(\mathbf{\lambda},\mathbf{y})^{m}\rangle^{1/m}_{\lambda} $. 

We simplify the stationarity equations in that we assume that parameters $q$ and $\Delta\chi$ are small near the transition to the glassy phase.  We then expand free energy $f_{1}$ from Eq.~\eqref{eq:FE-1RSB} to the overall fifth order in these parameters. An explicit expression for the expanded free energy is given in Appendix, Eq.~\eqref{eq:f1-asymptotic}. It allows us to derive explicit equations for the order parameters of the 1RSB solution. They are straightforward to derive but are rather lengthy. That is why we do not list them explicitly. We give here only the result of the asymptotic expansion in the small expansion parameter $\tau$. The results were again derived with the aid of program MATHEMATICA. The glassy phase of the Potts model allows for multiple solutions with discrete replica-symmetry breakings.  

We found a double degeneracy of the 1RSB solution. Both solutions have the same parameter $m$ to the two lowest orders in $\tau$
\begin{align}
m &\doteq \frac{p-2}{2}+\frac{36-12 p +p^2}{8(4-p)}\tau\ .
\end{align}
One non-trivial 1RSB solution then leads to order parameters
\begin{subequations}\label{eq:1RSB-Sol1}
\begin{align}
q^{(1)}&\doteq 0\ ,\\
\Delta\chi^{(1)}&\doteq \frac{2}{4-p}\tau + \frac{228-96 p +p^2}{6 (4-p)^3}\tau^2
\end{align}
\end{subequations}
while the second one has both parameters nonzero
\begin{subequations}\label{eq:1RSB-Sol2}
\begin{align}\label{eq:1RSB-Solq2}
q^{(2)}&\doteq \frac{-12+24 p -7 p^2}{3(4-p)^2 (p-2)}\tau^2 \ ,\\
\Delta\chi^{(2)}&\doteq \frac{2}{4-p}\tau - \frac{360-204 p -6 p^2+13 p^3}{6 (4-p)^3 (p-2)}\tau^2 \ . \label{eq:1RSB-Solchi2}
\end{align}\end{subequations}
Both the solution have the same asymptotic free energy to the fifth asymptotic order
\begin{widetext}
  \begin{equation}\label{f:1RSB:tau} \frac{\beta}{p-1} f_{1RSB}\doteq
    \frac{\tau ^3}{3 (4-p)} + \frac{(p (11 p-102)+204) \tau ^4}{12 (4-p)^3} 
    -\frac{(p (p ((18744-1103 p) p-120648)+325728)-317232) \tau ^5}{720
      (4-p)^5} \ . \end{equation}\end{widetext} 
Unlike the replica-symmetric solution we can see that the asymptotic expansion with small parameters $q$ and $\Delta\chi$ breaks down already at $p=4$ above which we expect a discontinuous transition from the paramagnetic to a 1RSB state at $T_{0} > 1$. 

The 1RSB solution has a higher free energy than the replica-symmetric one. The difference is of order $\tau^{3}$,  
\begin{equation}
f_{1RSB} - f_{RS} \doteq \frac {(p-2)^{2}(p-1) \tau^{3}}{3(4-p)(6-p)^{2}}\ .
\end{equation}
The two stationary states of the 1RSB free energy, Eq.~\eqref{eq:f1-asymptotic}, behave differently as a function of the parameter $p$. The former solution is physical for all values of $p$ unlike the latter that becomes unphysical for $p>p^{*}\approx 2.82$ where $q^{(2)}$ from Eq.~\eqref{eq:1RSB-Solq2} turns negative. It is also the region of the parameter $p$ where the first solution is locally stable as can be seen from the stability function      
\begin{multline} \label{eq:1RSB-instability1}
\Lambda^{(0)}_{1} = 
p - 1  - \beta^{2}\sum_{\alpha \beta}\left\langle \left\langle \rho\left[t^{\alpha\beta} - (1 - m)t^{\alpha}t^{\alpha} \right]^{2} \right\rangle_{\lambda} \right\rangle_{y}\\
\doteq\frac{\tau^{2}(p - 1)}{6(4 - p)^{2}} \left(
    7p^{2} - 24 p + 12\right) \ .
\end{multline} 
that is  positive in this region. We denoted
\begin{align}\label{eq:1RSB-t_alpha_beta}
t^{\alpha\beta}& = \frac{\sum_{A}e^{\alpha}_{A}e^{\beta}_{A} E_{A}(\mathbf{y},\mathbf{\lambda}) }{Z_{0}(\mathbf{y},\mathbf{\lambda}) }\ .
\end{align}

  
The first 1RSB solution $q^{(1)},\Delta\chi^{(1)}$ is not physically inconsistent and is locally stable for $p>p^{*}$. It should not break into solutions with a higher number of hierarchies of replicated spins. That is why it has been considered as the equilibrium state of the Potts glass on a ``high-temperature'' interval of the glassy phase. 

The two asymptotic solutions behave differently when approaching the Sherrington-Kirkpatrick model, $p\to 2$. The first is regular and goes over at $p=2$ to a new 1RSB solution with $q=0$.  Since both $\Delta\chi$ and $m$ are nontrivial, it is a solution breaking replica symmetry. This solution was not discussed in Ref.~\onlinecite{Janis06c} as we excluded $q=0$ case. The second solution is singular in the limit $p\to 2$. The singularity can be seen first the second order in the expansion parameter $\tau$. Consequently, the asymptotic limits $\tau \to 0$ and $(p-2) \to 0$ do not commute and the result depends on the value of ratio $\tau/(p-2)$. Due to this singularity we cannot unambiguously continue the second solution the  Sherrington-Kirkpatrick model.  This non-analyticity is connected with emergence of spin-reflection symmetry in the two-component Potts (Ising) model.       

\subsection{$K$RSB solution}
\label{sec:AS-KRSB}

Although we have a locally stable solution of the 1RSB free energy for $p>p^{*}$, there is another unstable solution that decays to a solution with a higher number of hierarchies of the replicated spins. We hence can investigate possible solutions of free energies with an arbitrary number of spin hierarchies. 

We start with stationarity equations for the $K$RSB free energy from Eq.~\eqref{eq:FE-K_general}. The equation for the order parameter $q$ reads
\begin{equation}\label{eq:KRSB-q}
(p-1)q=\langle\langle 
t^{\alpha}\rangle_{K} \langle 
t^{\alpha}\rangle_{K} \rangle_{y},
\end{equation}
where we denoted  $\langle F\rangle_{l}(y,\{\lambda\}_{l+1})= \int \mathcal{D}_{(p-1)}(\lambda_{l})\ \rho_{l}(y,\{\lambda\}_{l})\langle F\rangle_{l-1}(y,\{\{\lambda\}_{l}) $, $\rho_{l}(y,\{\lambda\}_{l}) = Z_{l-1}(y,\{\lambda\}_{l})^{m_{l}}/\langle Z_{l-1}^{m_{l}}\rangle_{\lambda_{l}}(y,\{\lambda\}_{l+1})$. We further abbreviated $\langle X\rangle_{\lambda_{l}} = \int \mathcal{D}_{(p-1)} \lambda_{l}\ X$. 
Equations for the overlap susceptibilities are 
\begin{align}\label{eq:KRSB-Dchi}
(p-1)\Delta\chi_{l}& 
= \langle\langle \langle 
t^{\alpha}\rangle_{l-1} \langle 
t^{\alpha}\rangle_{l-1}
\rangle_{K} \rangle_{y} \nonumber\\ & 
- \langle\langle \langle 
t^{\alpha}\rangle_{l} \langle 
t^{\alpha}\rangle_{l}
\rangle_{K} \rangle_{y} \ .
\end{align}
They are accompanied by equations for scaling parameters $m_{l}$ having the following form
\begin{align}\label{eq:KRSB-m}
m_{l}\frac{\beta^{2}}{4}(p-1)\Delta\chi_{l}&=\frac{\langle\langle \ln
Z_{l-1}\rangle_{K}\rangle_{y} - \langle\langle \ln
Z_{l}\rangle_{K}
\rangle_{y}}{2(q+\sum_{i=l+1}^{K}\Delta \chi_{i})+\Delta
\chi_{l}}
\end{align}
with $l=1,...,K$. 

One cannot solve these equations fully but we can reach their solution in the asymptotic limit    below the transition temperature to the glassy phase as done for the replica-symmetric and 1RSB solutions. We obtain only a single solution for $K > 1$ within the leading-order asymptotic limit $\tau\to 0$
\begin{align}\label{eq:KRSB-q_As}
q^{K}&\doteq -\frac{1}{3K^{2}}\frac{12-24p+7p^{2}}{(4-p)^{2}(p-2)}\tau^{2}\ ,\\
\label{eq:KRSB-Dchi_As}
\Delta\chi_{l}^{K}&\doteq \frac{1}{K}\frac{2}{(4-p)}\tau \ ,
\\  \label{sol:chi}
m_{l}^{K}& \doteq\frac{p-2}{2}+
\frac{2}{4-p}\left[3+\frac{3}{2}p-p^{2}\right. \nonumber \\ & \left. \qquad
+\left(3-6p+\frac{7}{4}p^{2}\right)\frac{2l-1}{2K}\right]\tau\ .
\end{align}
We can see that the $K$RSB solution behaves unphysically in the same way as the second 1RSB solution does. The averaged square of local magnetizations is negative for $p > p^{*}$ where the first 1RSB solution is locally stable. Negativity of $q$ means that local magnetizations are  imaginary and the solution is unphysical. This deficiency, however, decreases with the increasing number of spin hierarchies and disappears in the limit $K \to \infty$. It is analogous to negativity of entropy in the low-temperature solutions of $K$RSB approximations of the Sherrington-Kirkpatrick model.  

%
 
Free-energy density of the $K$RSB solution increases with increasing the number of hierarchical levels   
\begin{multline}\label{eq:KRSB-FE}
\frac{\beta}{p-1} f_{KRSB}\doteq \frac{ \tau ^3}{3 (p-4)} 
+ \frac{ (p (11 p-102)+204) \tau ^4}{12 (p-4)^3} \\ + \frac{ (p (p (p (16 p-265)+1686)-4532)+4408) \tau ^5}{10 (p-4)^5} \\ -\ \frac{\left(7 p^2-24 p+12\right)^2 \tau ^5}{720 K^4    (p-4)^5}
 \end{multline}
and reaches its maximum at $K=\infty$. Since the overlap susceptibilities are linear in $1/K$, the limit $K\to \infty$ leads to a solution with a continuous order-parameter function.

\subsection{Continuous replica-symmetry breaking}
\label{sec:AS-Continuous}

We showed in the preceding section that the Parisi continuous full RSB solution is not isolated and there is a cascade of discrete $K$RSB states, even in the region of stability of a 1RSB solution, converging to the full continuous RSB. The asymptotic expansions of these solutions are singular in the limit of the  Ising spin glass, $p=2$. The singularity, however, vanishes in the limit $K=\infty$ and the asymptotic expansion in $\tau$ of the Parisi solution appears to be analytic around $p=2$. The leading asymptotic order of the full RSB solution can then be obtained from the limit of the discrete $K$RSB approximations of the preceding section. To lift the degeneracy in the asymptotic region $T \to T_{c}$, we have to expand the order-parameter function to higher powers  of the small parameter $\tau$. We use the explicit representation in Eqs.~\eqref{eq:FE-continuous} to determine the asymptotic limit of the Parisi solution near the transition temperature. 

The characteristic function for the Parisi solution is a ''dynamical magnetization'' with the following representation via an evolution operator 
 \begin{multline}\label{eq:FRSB-q_prime} \qquad
   g^{\prime}_{\alpha}(\lambda,\mathbf{h}) =
  \mathbb{E}(X,\mathbf{h};\lambda,0)\circ g^{\prime}_{0,\alpha}
  \\ \equiv\mathbb{T}_{\nu} \exp\left\{X\int_0^{\lambda} d\nu
\left[\frac12
\partial_{\bar{h}^{\beta}}\partial_{\bar{h}^{\beta}} +
m(\nu) g^{\prime}_{\beta}(\nu, \textbf{h} +
\bar{\textbf{h}})\partial_{\bar{h}^{\beta}} \right]
  \right\}\\ \frac{\partial g_{0}(\textbf{h}+ \bar{\textbf{h}})}{\partial h^{\alpha}}\bigg|_{\bar{\textbf{h}} = 0}\ .
\end{multline}

The defining equation for the stationary value of parameter $q$ reads 
 \begin{equation}\label{eq:FRSB-q}
\beta^{2}(p-1)q=\langle
g^{\prime}_{\alpha}(1,\mathbf{h}_{\eta})g^{\prime}_{\alpha}(1,\mathbf{h}_{\eta})\rangle_{\eta}
\end{equation}
where $\mathbf{h}_{\eta}\equiv\mathbf{h}+\eta\sqrt{q}$. Parameter $X$ is determined from
\begin{multline}\label{eq:FRSB-X}
\beta^{2}(p-1)X=\langle
\mathbb{E}(X,\mathbf{h}_{\eta};1,0)\circ \left[g^{\prime}_{0,\alpha}(\mathbf{h}_{\eta})g^{\prime}_{0,\alpha}(\mathbf{h}_{\eta})\right]\rangle_{\eta} \\
-\langle g^{\prime}_{\alpha}(1,\mathbf{h}_{\eta})g^{\prime}_{\alpha}(1,\mathbf{h}_{\eta})\rangle_{\eta} \ .
\end{multline}
Order-parameter function $m(\lambda)$ is obtained from an identity 
\begin{multline}\label{eq:FRSB-m}
\beta^{2}(p-1)X \lambda=\langle
\mathbb{E}(X,\mathbf{h}_{\eta};1,0)\circ\left[g^{\prime}_{0,\alpha}(\mathbf{h}_{\eta})g^{\prime}_{0,\alpha}(\mathbf{h}_{\eta})\right]\rangle_{\eta}
\\-\langle
\mathbb{E}(X,\mathbf{h}_{\eta};1,\lambda)\circ\left[g^{\prime}_{\alpha}(\lambda,\mathbf{h}_{\eta})g^{\prime}_{\alpha}(\lambda,\mathbf{h}_{\eta})\right]\rangle_{\eta}
\end{multline}
valid for $\lambda \in (0,1)$.  

We next use the knowledge from the limit $K\to\infty$ of the discrete $K$RSB solutions, namely 
\begin{equation}\label{eq:FRSB-q0}
g^{\prime}_{\alpha}(1,0)=0
\end{equation}
expressing consistency of the continuous limit.  

Equation \eqref{eq:FRSB-m} holds for all index variables $\lambda\in (0,1)$. Its form is unsuitable for determination of order-parameter function $m(\lambda)$. We can, however, perform analytic operations on both sides of this equation so that to transform it to a more suitable form. Applying derivative with respect to $\lambda$ leads to a condition of marginal stability.\cite{Janis08a} It reads  
\begin{equation}\label{eq:marg-cond}
\beta^{2}(p-1) =\langle
\mathbb{E}(X,0;1,\lambda)\circ\left[g^{\prime\prime}_{\alpha\beta}(\lambda,0)g^{\prime\prime}_{\alpha\beta}(\lambda,0)\right]\ .
\end{equation}
To derive this form we used representation \eqref{eq:FE-continuous} and differential equation \eqref{eq:g0-DE}. Further derivative of the above equation with respect to $\lambda$ leads to an explicit representation for the order-parameter function%
\begin{equation}\label{eq:m_d2}
2 m(\lambda) = \frac{\mathbb{E}(X, 0;1,\lambda)\circ\left[g^{\prime\prime\prime}_{\alpha\beta\gamma}(\lambda)g^{\prime\prime\prime}_{\alpha\beta\gamma}(\lambda)\right]}{\mathbb{E}(X, 0;1,\lambda)\circ\left[g^{\prime\prime}_{\alpha\beta}(\lambda)g^{\prime\prime}_{\beta\gamma}(\lambda)g^{\prime\prime}_{\gamma\alpha}(\lambda)\right]} \ .
\end{equation}
Evaluating this expression at the transition point $T_{c}= 1$ where $X=0$ and $\mathbb{E} = 1$ we obtain $2 m(0) = p-2$. Below the critical temperature where $X>0$ we expand order-parameter function $m(\lambda)$ in $\lambda X$ and $X$ as independent parameters 
\begin{align}\label{eq:FRSB-m_exp}
m(\lambda)& =  \sum_{j=0,k=j}m[j,k]\lambda^{j}X^{k}
\end{align}
where $k$ is the order of the asymptotic expansion and determines the asymptotic precision. We expand analogously the $\lambda$-dependent free energy so that to be able to resolve evolution operator $\mathbb{E}$. We have to keep dependence of free energy on an external magnetic field and hence  
\begin{align}
\label{eq:FRSB-g_exp}
g(\lambda;\mathbf{h}) & = g_{0}(\mathbf{h}) +  \sum_{j=1,k=j}g[j,k;\mathbf{h}]\lambda^{j}X^{k}\ .
\end{align}

We expand all quantities in powers of $X$ and $\lambda X$ and solve each equation for individual orders independently. We do not list here all equations for the expansion parameters. Nontrivial expansion parameters $m[j,k]$ and $g(j,k) = g[j,k;0]$ to the order $k=5$ are listed in Appendix.  

The low-temperature glassy phase is reached when there is a nontrivial solution of Eq.~\eqref{eq:FRSB-X} for parameter $X$. In the asymptotic limit $X\to 0$ we obtain an equation allowing for a nontrivial solution if $\beta > 1$
\begin{widetext}
\begin{multline}\label{eq:FRSB-X_var}
0 = \frac{1}{6} X^3
   \beta ^8 \left(\beta ^2 \left(4 p^2 (3 m[0,0]-19)+6 p \left(m[0,0]^2-29
   m[0,0]+114\right)\right.\right.
   \\\left.\left.-12 \left(2 m[0,0]^2-33 m[0,0]+105\right)+p^3\right)+2 (p-4)
   (3 m[0,1]+m[1,1])\right)\\
   +\frac{1}{2} X^2 \beta ^8 \left(2 p (m[0,0]-10)-8 m[0,0]+p^2+50\right)+(p-4) X \beta ^6+\beta^2(\beta^2-1)\ .
\end{multline}
Using expansion coefficients $m[j,k]$ from Appendix we obtain the first four exact powers in parameter $\tau = (\beta - 1)/\beta $
\begin{multline}\label{eq:FRSB-Xtau}
X=\frac{2}{4-p}\tau 
-\frac{(p (p+12)-36) }{(4-p)^3}\tau ^2 
-\frac{(560-4 p (p (11 p-57)+136)) }{(4-p)^5}\tau^3 \\
-\frac{(p (p (p (p (71 p-705)+5832)-25832)+54960)-44560)
}{(4-p)^7}\tau^4+O(\tau^{5}) 
\end{multline}
being small (finite) for $p<4$.

We can analogously evaluate free energy of the Potts glass. Using again expansion coefficients $m[j,k]$ we obtain an asymptotic expansion up to the fifth order in $X$ of $g(1,0)$
\begin{multline}\label{g1tau}
\frac{g(1,0)}{p-1}=\frac{\log (p)}{p-1} \\ +\frac{X \beta
^2}{2}+\frac{1}{8} (p-4) X^2 \beta ^4 
-\frac{1}{48} (p (p+38)-112) X^3 \beta ^6 
-\frac{1}{384} (p (p (19
   p+194)-3828)+7896) X^4 \beta ^8
\\
+\frac{X^5 \beta ^4 \left(4 m[3,3]+3 (p (p ((1765-576 p)
p+45394)-322428)+493032) \beta ^6\right)}{5760}
+O(X^{6})\ .
\end{multline}
\end{widetext}
We insert the asymptotic values of parameters $X$ and $m[3,3]$ and obtain an explicit dependence on the small parameter $\tau$
\begin{multline}\label{eq:ftau}
  \frac{\beta}{p-1} f_{c}(\tau)\doteq \frac{1}{3 (4-p)} \tau ^3
  +\frac{(p (11 p-102)+204) }{12 (4-p)^3}\tau ^4 \\ +\frac{(p (p (p
    (16 p-265)+1686)-4532)+4408) }{10 (4-p)^5}\tau ^5\ .
\end{multline}
It is easy to demonstrate that free energy of the full RSB solution is higher than free energies of the discrete RSB solutions. We have  
\begin{equation}\label{eq:fdiff} \beta (f_{c}-f_{KRSB})\doteq
  \frac{(p-1) (p (7 p-24)+12)^2 \tau ^5}{720 K^{4}(4-p)^5} 
\end{equation} 
and 
\begin{equation}
  \beta(f_{c } - f_{RS}) \doteq \frac{(p - 1) (p - 2)^{2}\tau^{3}}{3(4 - p)(6
    - p)^{2}}\ .
\end{equation} 
Parisi-like solution with a continuous order-parameter function has the highest free energy as in the SK model and represents the true thermodynamic equilibrium of the Potts glass for $2\le p < 4$.

\section{Discussion and conclusions}
\label{sec:DC}

We studied in this paper the asymptotic behavior of the mean-field $p$-state Potts glass below the transition temperature to the glassy phase defined by instability of the replica-symmetric solution. We analyzed separately the replica-symmetric solution, solutions with 1RSB, $K$RSB for $K>1$, and a solution with continuous RSB. We separated the 1RSB scheme, since its free energy is the only one with two distinct stationary states. We found that the RSB solutions peel off from the replica-symmetric one continuously for $p\le 4$ so that the small expansion parameters are $q$, the averaged square of the local magnetization, and the overlap susceptibilities $\Delta\chi_{l}$ between neighboring hierarchies of spin variables $l-1$ and $l$. We expanded these order parameters to the fifth order in $\tau = 1 - T/T_{c}$ to distinguish individual states. The asymptotic expansion allowed us to analyze the behavior of the Potts glass also as a function of parameter $p$ as a continuous variable connecting solutions for $2\le p < 4$.   We were able to distinguish two regions where the solutions with finite-many replica hierarchies behave differently. The two regions are separated by a critical value  $p= p^{*}\approx 2.82$.  

The  Potts model in the region with $2\le p <p^{*}$ has only a single solution representing the true equilibrium. It is the solution with continuous RSB being marginally stable.  The solutions for the schemes with finite-many replica hierarchies are all unstable as in the case of the Ising spin glass. The $K$RSB solutions are, however, non-analytic around the Ising limit, $p=2$. A singularity in the asymptotic expansion of these solutions emerges in the second order of $\tau$ for $p=2$.  It means that there is no analytic continuation of the $K$RSB solutions from the Ising spin glass to the Potts one. This non-analyticity is caused by spin-reflection symmetry present only in the Ising model.  The 1RSB free energy of the Potts model deserves special attention. Apart from the asymptotic solution with $q\neq 0$ singular at $p=2$  we found another solution  with $q= 0$ free of any singularity in the limit $p\to 2$. This solution can hence be analytically continued to $p=2$ to a 1RSB solution with non-zero parameters $\Delta\chi$ and $m$. Such analytic continuation has a lower free energy than the standard 1RSB solution with $q>0$ discussed in Ref.~\onlinecite{Janis06c}. It is interesting to note that the 1RSB solution of the SK model  with $q=0$ decays to solutions with a higher number of replica hierarchies. The asymptotic expansion of parameters $\Delta\chi$ and $m$ for the general $K$RSB solution reads
\begin{align*}
\Delta\chi^K_j &= \frac{1}{K}\tau + \frac{(6 K^2-1)}{6 K^3} \tau ^2\ ,\\ 
m^{K}_{j} &= \frac{(2(K-j+1)-1)}{K}\tau \\ &\qquad + \frac{(12K^{2}+1)(2(K-j+1)-1)}{6K^3}\tau^2\ .
\end{align*}
All these $K$RSB solutions with $q=0$ are unstable, have a lower free energy than the $K$RSB solutions with $q>0$ and converge towards the Parisi full RSB solution where $q=0$ as well. 

The RSB schemes with finite-many hierarchies change their behavior after passing a critical value $p=p^{*}$ where the stability function of the first 1RSB solution, $\Lambda^{(0)}_{1}$ becomes positive, indicating its local stability. Moreover, the second 1RSB solution and all other $K$RSB schemes turn unphysical, since the averaged square of the local magnetization goes through zero and gets negative, $q<0$ for $p>p^{*}$. It is, however, important that negativity of parameter $q$ decreases with increasing $K$ and approaches zero in the continuous limit $K\to \infty$.  The unphysical $K$RSB states converge then towards a solution with full continuous RSB for which $q=0$. Hence, the continuous RSB solution does not experience any unphysical behavior for $p\ge p^{*}$. It remains marginally stable for all values of $p$, has the highest free energy from all studied construction schemes, and is thermodynamically homogeneous. The only observable change in the full RSB solution for $p>p^{*}$, as discussed in Ref. \onlinecite{Janis11a}, is a change in the sign of the derivative of order-parameter function $m(\lambda)$, cf. coefficient $m[1,1]$ in Appendix. 

To conclude, we analyzed the asymptotic behavior of the mean-field random Potts glass with the number of states $p< 4$ below the transition to the glassy phase.  We demonstrated that a Parisi-like solution with continuous replica-symmetry breaking emerges simultaneously with the instability of the replica-symmetric solution. Its existence is independent of local stability of the 1RSB solution observed for $p> p^{*}\approx 2.82$. We found that the solution with continuous RSB in the Potts model is a limit of other unstable solutions with discrete RSB that decay towards it alike in the Sherrington-Kirkpatrick model. These solutions are unphysical, since the averaged squared local magnetization is negative, but the resulting limit with continuous RSB is free of any unphysical behavior.  We studied in this paper only the Potts model with $p\le 4$, but our construction of  the asymptotic expansion in parameter $X= q_{EA} - q_{SK} $ can be extended also to $p =4 + \epsilon$ for which $X\ll 1$. There we expect that replica-symmetry breaking solutions emerge above the critical temperature of instability of the replica-symmetric one. The solution with continuous RSB does not seem to display any singularity at $p=4$.  The question to be answered in the Potts model with $p>4$ is: which solution represents the true equilibrium state in the region of coexistence of the replica-symmetric and replica-symmetry breaking solutions?                                
%

Research on this problem was carried out within project AV0Z10100520
of the Academy of Sciences of the Czech Republic.

\appendix
\section{Expanded free energies near the critical transition temperature to the glassy phase}
\label{sec:App}

Free energy of the replica-symmetric solution has only a single parameter $q$ that we can use as the expansion parameter below the critical transition temperature to the glassy phase. We expand free energy to the fifth order, although third order would be sufficient to distinguish it from the replica-symmetry breaking ones.  We obtain for $q\ll 1$
\begin{widetext}
\begin{multline}\label{eq:FE-RS_beta}
\frac{\beta}{p-1} f_{RS}\doteq\frac{\beta^2}{4} q^2 \left(\beta^2 -1\right) +\frac{\beta^6}{12}
   (p-6) q^3 +\frac{\beta^8}{48} \left(p^2-30 p+90\right) q^4 +\frac{\beta^{10}}{240} \left(p^3-114 p^2+1236 p-2520\right) q^5\ .
\end{multline}

We explicitly expand only 1RSB  from all discrete $K$RSB solutions where we use two small expansion parameters $q$ and $\Delta\chi$. It is necessary to expand it to the fifth order so that to distinguish it from higher-order RSB solutions and the continuous full RSB one. We obtain
\begin{multline}\label{eq:f1-asymptotic}
\frac{\beta}{p-1}f_{1}\doteq \frac{\beta}{p-1}f_{RS}
    +\frac{1-m}{2}\left( q \left(\beta ^2-1\right) \beta ^2+\frac{1}{2} (p-6) q^2 \beta ^6\right.
  \\\left.
  +\frac{1}{6} \left(p^2-30 p+90\right) q^3 \beta ^8+\frac{1}{24} \left(p^3-114 p^2+1236 p-2520\right) q^4
   \beta ^{10}\right) \Delta\chi \\
   +\frac{1-m}{4} \left(\left(\beta ^2-1\right) \beta ^2+q \beta ^6 (2
   m+p-6)+\frac{1}{2} q^2 \beta ^8 \left(10 m (p-4)+p^2-30 p+90\right)
   \right.\\ \left.+\frac{1}{6} q^3 \beta
   ^{10} \left(2 m \left(19 p^2-276 p+630\right)+p^3-114 p^2+1236 p-2520\right)\right)\Delta\chi^2
    \\  +\frac{1-m}{6} \left(m \beta ^6+\frac{1}{2}
   (p-6) \beta ^6+\frac{1}{2} q \beta ^8 \left(6 m^2+6 m (2 p-9)+p^2-30 p+90\right)
      \right.\\
      \left. +\frac{1}{4} q^2 \beta ^{10} \left(30
   m^2 (3 p-10)+6 m \left(8 p^2-129 p+308\right)+p^3-114 p^2+1236 p-2520\right)\right)\Delta\chi^3
    \\  +\frac{1-m}{8}\left(
   m^2 \beta ^8+\frac{1}{6} m (12 p-54) \beta ^8+\frac{1}{6}
   \left(p^2-30 p+90\right) \beta ^8
\right.\\
\left. +q \left(4 m^3 \beta ^{10}+4 m^2 (5 p-19) \beta ^{10}+\frac{1}{6} m \beta ^8 \left(50 p^2
   \beta ^2-840 p \beta ^2+2064 \beta ^2\right)\right.\right.\\
   \left.\left.+\frac{1}{6} \beta ^8 \left(p^3 \beta ^2-114 p^2 \beta
   ^2+1236 p \beta ^2-2520 \beta ^2\right)\right)\right)\Delta\chi^4
    \\  +\frac{1-m}{10}\left(
   m^3 \beta ^{10}+m^2 (5 p-19) \beta ^{10}+\frac{1}{24} m \left(50 p^2-840 p+2064\right) \beta
   ^{10}\right.\\
   \left.+\frac{1}{24} \left(p^3-114 p^2+1236 p-2520\right) \beta ^{10}\right)\Delta\chi^5\ .
\end{multline}
\end{widetext}

The full continuous free energy is defined via its expansion coefficients in Eq.\eqref{eq:FRSB-g_exp}. Their values to the fifth order at zero magnetic field read
\begin{align}\label{eq:sol-g}
g (1,1) & = \frac{\beta ^2}{2}\ ,\\
g (1,l) & = 0, \qquad l > 1\ ,\\
g (2,2) & = \frac{1}{4} \beta ^4 (m [0,0]-1)\ ,\\
g (2,3) & = \frac{1}{4} \beta ^4 m [0,1]\ ,\\
g (3,3) & = \frac{1}{12} \beta ^4 \left(\beta ^2 (m [0,0]-1) (2 m [0,0]+p-6)\right. \nonumber \\  & \left.  +m [1,1]\right)\ ,
\end{align}
\begin{align}
g (2,4) & = \frac{1}{8} \beta ^4 m [0,2]\ ,\\
g (3,4)  & = \frac{1}{24} \beta ^4 \left(2 \beta ^2 m [0,1] (4 m [0,0]+p-8)\right. \nonumber \\  &\left.  +m (1,2)\right)\ ,\\
g (4,4) & = \frac{1}{96} \beta ^4 \left(2 \beta ^4 (m [0,0]-1) \left(6 p (2 m [0,0]-5)\right. \right. \nonumber \\ & \left. \left. +6 \left(m [0,0]^2  -9 m [0,0]+15\right)+p^2\right)\right. \nonumber \\ & \left. +2 \beta ^2 m [1,1] (6 m [0,0]+p-10)+m [2,2]\right)\ ,\\
g (2,5) & = \frac{1}{24} \beta ^4 m [0,3]\ ,\\
g (3,5) & = \frac{1}{72} \beta ^4 \left(3 \beta ^2 \left(m [0,2] (4 m [0,0]+p-8)\right. \right.\nonumber \\ & \left. \left. +4 m [0,1]^2\right)+m [1,3]\right)\ , \\
g (4,5) & =\frac{1}{288} \beta ^4 \left(6 \beta ^4 m [0,1] \left(6 p (4 m [0,0]-7)\right. \right. \nonumber \\ & \left. \left. +6 \left(3 m [0,0]^2-20 m [0,0]+24\right)+p^2\right)\right. \nonumber \\  & \left. +3 \beta ^2 (m [1,2] (6 m [0,0]+p-10)\right. \nonumber \\ & \left.  +12 m [0,1] m [1,1])+m [2,3]\right) \\
g (5,5) & = \frac{1}{1440}\beta ^4 \left(6 \beta ^4 m [1,1] \left(p (38 m [0,0]-56)\right. \right. \nonumber \\ & \left. \left. +36 m [0,0]^2-206 m [0,0] +p^2+212\right)  \right. \nonumber \\ & \left. + 6 \beta ^6 (m [0,0]-1) \left(2 p^2 (25 m [0,0]-57) \right.\right. \nonumber \\ & \left.\left.  +12 p \left(10 m [0,0]^2-70 m [0,0]+103\right) \right.\right. \nonumber \\ & \left. \left. +24 \left(m [0,0]^3-19 m [0,0]^2+86 m [0,0]\right.\right.\right. \nonumber \\ & \left. \left. \left. -105\right)+p^3\right) +3 \beta ^2 \left(m [2,2] (8 m [0,0]+p\right. \right. \nonumber \\ & \left. \left.-12)+12 m [1,1]^2\right)+m [3,3]\right) \ .
\end{align}

The corresponding expansion coefficients of the order-parameter function to the third order are 
\begin{align}\label{eq:sol-m}
 m[0,0]& = \frac{p-2}{2}\ , \\
 m[0,1]& = \frac{1}{2} ((3-2 p) p+6) \beta ^2\ , \\
 m[1,1]& = \frac{1}{4} (p (7 p-24)+12) \beta ^2\ , \\
 m[0,2] & = (p+2) (p (p+9)-27) \beta ^4\ , \\
 m[1,2] & = -\frac{1}{2} (p (p (17 p+19)-228)+204) \beta ^4\ , \\
 m[2,2] & = \frac{3}{4} (p-2) (p (25 p-32)-44) \beta ^4\ ,\\
 m[0,3] & = -\frac{3}{2} (p (p ((p-29) p+332)+68)\nonumber \\ & -1504) \beta ^6\ , \\
 m[1,3] & = \frac{3}{2} (p (p (2 p (5 p+46)+935)-4296)\nonumber \\ &+3684) \beta ^6 \ ,\\
 m[2,3] & = -\frac{9}{4} (p (p (p (52 p+285)-1006)-1068)\nonumber \\ & +3016) \beta ^6\ ,\\
m[3,3] & = \frac{9}{8} \beta ^6 \left(p (p (p (283 p-488)-1208)+1184)\right. \nonumber \\ & \left.   +1776 \right)
 \ .
\end{align}

 \end{document}